\documentclass[superscriptaddress,reprint,amsmath,amssymb,pra]{revtex4-1}
\usepackage{hyperref}
\usepackage{physics}
\usepackage{bm}
\usepackage{ulem} 
\usepackage{graphicx}

\begin{document}

\newcommand{\edo}{\textcolor{teal}}
\newcommand{\uday}{\textcolor{violet}}
\newcommand{\john}{\textcolor{blue}}
\newcommand{\qqedit}{\textcolor{magenta}}
\newcommand{\james}{\textcolor{red}}
\newcommand{\jake}{\textcolor{green}}

\title{C-band single photons from a trapped ion via two-stage frequency conversion}
\author{J. Hannegan}
\affiliation{Institute for Research in Electronics and Applied Physics, University of Maryland, College Park, Maryland 20742, USA}
\affiliation{Department of Physics, University of Maryland, College Park, MD, 20742, USA}
\author{U. Saha}
\affiliation{Institute for Research in Electronics and Applied Physics, University of Maryland, College Park, Maryland 20742, USA}
\affiliation{Department of Electrical and Computer Engineering, University of Maryland, College Park, MD, 20742, USA}
\author{J. D. Siverns}
\affiliation{Institute for Research in Electronics and Applied Physics, University of Maryland, College Park, Maryland 20742, USA}
\affiliation{Department of Physics, University of Maryland, College Park, MD, 20742, USA}
\author{J. Cassell}
\affiliation{Institute for Research in Electronics and Applied Physics, University of Maryland, College Park, Maryland 20742, USA}
\affiliation{Department of Physics, University of Maryland, College Park, MD, 20742, USA}
\author{E. Waks}
\affiliation{Institute for Research in Electronics and Applied Physics, University of Maryland, College Park, Maryland 20742, USA}
\affiliation{Department of Electrical and Computer Engineering, University of Maryland, College Park, MD, 20742, USA}
\affiliation{Department of Physics, University of Maryland, College Park, MD, 20742, USA}
\author{Q. Quraishi}
\affiliation{United States Army Research Laboratory, 2800 Powder Mill Road, Adelphi, Maryland 20783, USA}
\affiliation{Institute for Research in Electronics and Applied Physics, University of Maryland, College Park, Maryland 20742, USA}
\affiliation{Department of Physics, University of Maryland, College Park, MD, 20742, USA}



\begin{abstract}
Fiber-based quantum networks require photons at telecommunications wavelengths to interconnect qubits separated by long distances.
Trapped ions are leading candidates for quantum networking with high-fidelity two-qubit gates, long coherence times, and the ability to readily emit photons entangled with the ion's internal qubit states. 
However, trapped ions typically emit photons at wavelengths incompatible with telecommunications fiber. Here, we demonstrate frequency conversion of visible photons, emitted from the S-P dipole transition of a trapped Ba+ ion, into the telecommunications C-band. 
These results are an important step towards enabling a long-distance trapped ion quantum internet. 

\end{abstract}
\maketitle
Quantum computing \cite{wright2019,erhard2019,pino2020}, simulation \cite{kokail2019,Davoudi2020}, and communication platforms based on trapped ions \cite{stephenson2020high,hucul2015modular} are at the forefront of quantum information science. 
Trapped ion systems are well suited for quantum networking, given their long coherence times~\cite{wang2021single}, high single~\cite{harty2014} and two-qubit~\cite{Gaebler2016,ballance2016high,srinivas2021high} gate fidelities, and their ability to emit photons entangled with the ion's internal states~\cite{Siverns17}. 
Recently, an elementary ion-based network demonstrated entanglement rates above 180 Hz with fidelities of 0.94, the highest rates of any system with a fidelity above 0.9~\cite{stephenson2020high}, but with a node separation of only two meters.

Trapped ion quantum networks have been limited in range due to photon emission at UV and visible wavelengths, where light suffers large fiber-optic propagation losses~\cite{hucul2015modular,inlek2017multispecies,stephenson2020high,Bock2018,Walker2018,Krutyanskiy2019}.
Additionally, this prevents the integration of these networks into existing telecommunications infrastructure.
Of particular interest are photons produced via S-P dipole transitions, enabling direct entanglement between the photons and commonly used ground state qubits of ions such as
Yb$^{+}$~\cite{moehring2007entanglement,kobel2021deterministic}, Ba$^{+}$~\cite{auchter2014,Siverns17,crocker2019high}, and Sr$^{+}$~\cite{stephenson2020high}.
Ground-state qubits currently demonstrate the longest coherence times in trapped ions~\cite{wang2021single}, as well as leading two-qubit gate fidelities~\cite{Gaebler2016,ballance2016high,srinivas2021high}.

To date, quantum frequency conversion (QFC) has been used to down-convert ion-emitted photons in the near-IR ($\approx850$ nm) to telecom wavelengths~\cite{Bock2018,Krutyanskiy2019,Walker2018}, greatly improving potential networking range. The largest ion-photon and ion-ion entanglement rates, however, have been demonstrated in systems emitting UV or visible photons \cite{stephenson2020high,hucul2015modular}. 
QFC of visible ($493$ nm) ion-emitted photons, which may be directly entangled with ground state ion qubits, to the near IR ($\approx 780$ nm)~\cite{Siverns19convert,Siverns19slow} has been demonstrated, serving to both increase networking range and providing a pathway for the implementation of hybrid networks using both trapped ion and neutral atomic systems~\cite{Siverns19slow,Craddock19,hanneganSiverns2021}.

In this work, we demonstrate QFC of S-P dipole generated 493-nm photons originating from a single trapped $^{138}$Ba$^+$ ion to the telecommunications C-band.
To span this large frequency range, we employ two concatenated QFC stages.
This two-step approach circumvents the noise produced via spontaneous parametric down conversion (SPDC) present in any single-stage QFC scheme taking visible light to the C-band~\cite{Krutyanskiy2019,pelc2011long}.
Additionally, this scheme offers the advantage of having the first-stage conversion resonant with neutral-atom based systems for potential integration into a hybrid quantum network~\cite{Craddock19,Siverns19slow,hanneganSiverns2021,maring2017photonic}.
We demonstrate quantum efficiencies of 0.37 and 0.15 for the first- and second-stage respectively, using low-level laser light.
We show that each QFC step preserves both the pulse shape and quantum statistics of 493-nm photons produced by the Ba$^+$ ion.
These results demonstrate the largest QFC shifts achieved using trapped ions, and provide a key step towards long-distance hybrid quantum networks using trapped ion ground-state qubits. 

\begin{figure*}[htbp]
\centering\includegraphics[width=1\textwidth]{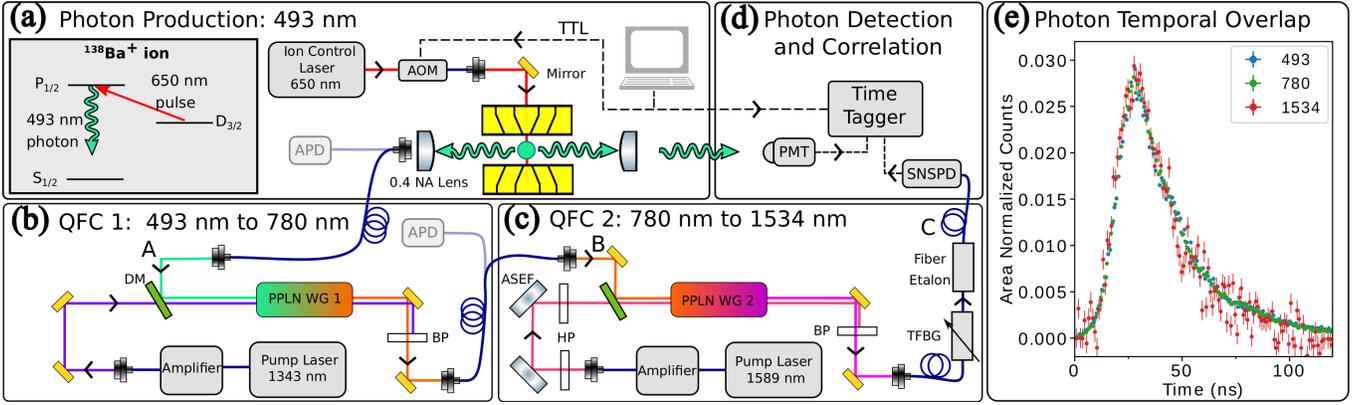}
\caption{(a) A TTL pulse triggers 650-nm light, exciting a $^{138}$Ba$^+$ ion and allowing for the spontaneous emission of a 493-nm photon. Emitted photons are sent to either a PMT or are fiber coupled and sent to an APD or QFC 1.
(b) QFC 1 combines 493-nm single photons with 1343-nm light in PPLN WG 1, producing fiber-coupled 780-nm single photons via DFG. After filtering via a 3 nm band pass filter (BP), these photons are either detected by an APD or sent to QFC 2.
(c) QFC 2 combines 780-nm single photons with 1589-nm light in PPLN WG 2, producing fiber-coupled 1534-nm single photons via DFG. A tunable fiber Bragg grating (TFBG) and fiber etalon frequency filter the output of QFC 2, reducing noise. To filter out 1534-nm light present in the pump, we use amplified spontaneous emission filters (ASEF) and 1550 nm high pass filters (HP).
(d) Detection events for the PMT and SNSPD and TTL trigger signal are time-tagged for two-photon correlation measurements. (e) Overlap of single photon arrival time histograms at each wavelength. \label{fig:setup}}
\end{figure*}

We trap a single $^{138}$Ba$^+$ ion as described in \cite{Siverns19convert}.
A magnetic field defines the quantization axis perpendicular to two opposing 0.4 NA photon collection lenses (Fig \ref{fig:setup} (a)).
One lens collects single photons emitted by the ion and sends them to a free-space single photon counting photomultiplier tube (PMT).
We refer to the photons collected by this lens as PMT-photons.
The second lens couples single photons into a single-mode fiber coupling setup. 
These 493-nm  photons are either sent to a fiber-coupled single-photon counting avalanche photodiode (APD) or to QFC 1 and QFC 2 for down conversion (Figs \ref{fig:setup} (b) and (c)). 

To produce single photons from the ion, we use a three step process of initialization, excitation, and spontaneous emission \cite{Siverns17}. 
After a period of Doppler cooling, the ion is initialized into $\ket{5D_{3/2}, m_j=3/2}$ using a 781 ns pulse containing $\pi$ and ${\sigma}^-$ polarized 650-nm light and $\pi$ polarized 493-nm light.
After a 200 ns delay with no light incident on the ion, a 200 ns pulse of ${\sigma}^+$ polarized 650-nm light excites the ion to $\ket{6P_{1/2}, m_{j}=1/2}$. The ion can then decay to the $6S_{1/2}$ ground state manifold, emitting a 493-nm photon (Fig.~\ref{fig:setup}a).
This scheme can be used to generate entanglement between the polarization of the emitted photon and the spin of the ion ground state \cite{Siverns17}.
 
\begin{figure}[t]
\centering\includegraphics[width=1.0\columnwidth]{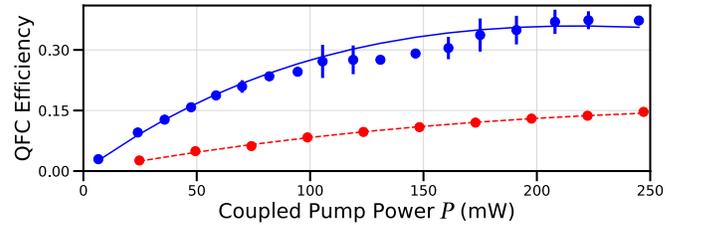}
\caption{Efficiencies of QFC 1 (blue) and QFC2 (red), measured using low level laser light. 
Lines represent fits to $\eta sin^2[(\pi/2)\sqrt{P/P_{m}}]$ where $P_{m}$ is the pump power at peak conversion efficiency, $\eta$.
Values of $\eta$ are 0.36 and 0.15 for QFC 1 and QFC 2 respectively. 
Uncertainties represent fluctuations in efficiency observed over measurement.}\label{fig:dfg}
\end{figure}

We convert 493-nm photons emitted by the ion to the C-band using two concatenated QFC setups, illustrated in Figs \ref{fig:setup}(b) and (c).
In QFC 1, 1343-nm light from a high intensity pump laser and 493-nm light, originating either from the ion or a laser used for testing and alignment, are coupled into a buried reverse-proton-exchange-fabricated waveguide in periodically poled lithium niobate (PPLN) (SRICO SR050117.4.C4).
Through difference frequency generation (DFG), the 493-nm light is converted to 780 nm and then coupled into an optical fiber, which is connected to either the APD or to QFC 2. 
We use optical filters (Semrock LL01-780 and FF01-1326/SP) at the output to remove 1343-nm pump light and 780-nm noise produced through anti-Stokes Raman processes in the PPLN waveguide. 
Using laser light, we measure the efficiency of QFC 1 from the output of the 493-nm input fiber to the output of the 780-nm fiber (A to B in Fig. \ref{fig:setup}) as shown in Fig. \ref{fig:dfg}.
At the peak QFC efficiency of $\approx$ 0.37, 320 cps noise are detected on our APD ($\approx 54\%$ detection efficiency) after the output 780-nm fiber, including $\approx100$ cps detector dark noise.

In QFC 2, 780-nm light from QFC 1 and 1589-nm pump light are coupled into a ridge waveguide fabricated in PPLN with a poling period of 18.75 $\mu$m (NTT-Photonics WD-1530-000-A-C-C-TEC). 
Difference frequency generation results in conversion of 780-nm photons to 1534 nm.
As with the first conversion stage, free-space interference filters (Semrock NIR01-1535/3) filter out excess pump signal and reduce Raman noise produced by the pump.
With the pump wavelength close to the target wavelength of 1534 nm, additional pump filtering is used in this stage, including a tunable fiber bragg grating (Advanced Photonics International, 19 GHz bandwidth), and a temperature-stabilized fiber etalon (Micron Optics FF24Z2 46.1 MHz bandwidth) at the output of the converter.
An amplified spontaneous emission filter (Coherent NoiseBlock) is used before the PPLN to remove any 1534-nm light present in the 1589-nm pump input.
The 493-nm photon's frequency is set by the ion's emission profile so the 1343-nm laser frequency determines the frequency of the 780-nm first-stage single-photon output.
The 1343-nm pump frequency is set to achieve simultaneous phase matching for both QFC stages, and to ensure that the converted 1534-nm light is transmitted through the final filtering stage.
The efficiency of this stage is measured with laser light as the ratio of the number of 1534-nm photons after filtering to the number of 780-nm photons at the output of the 780-nm fiber (B to C in Fig.~\ref{fig:setup} (c)).
After QFC 2, photons are detected using a superconducting nanowire single photon detector (SNSPD) ($\approx 78\%$ detection efficiency).
At a QFC efficiency of $\approx$ 0.15, shown in Fig.~\ref{fig:dfg}, we measure, $\approx 2950$ cps noise on average after filtering, with a filter transmission of $\approx 26\%$ for laser light, corresponding to a transmission of $\approx 19.7\%$ for ion-produced photons (linewidth $\Gamma/2\pi=14.8$ MHz \cite{Arnold19}), reducing the overall QFC efficiency to $\approx0.03$.

\begin{figure}[htbp]
\centering\includegraphics[width=1\columnwidth]{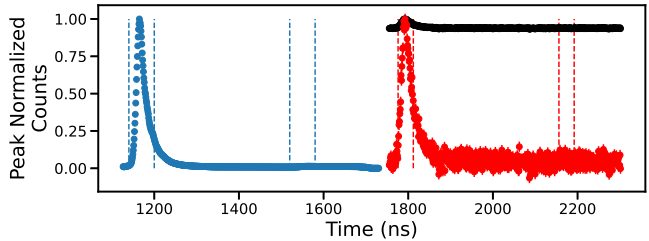}
\caption{Histograms of 493-nm PMT (blue) and 1534-nm (black) photon arrival times following the trigger TTL pulse. 
Dashed vertical lines indicate gate windows used for correlations (left) and noise measurements (right) for each curve. 
The red trace shows the background-subtracted 1534-nm data.
}\label{fig:photonShapes}
\end{figure}

Additional non-fundamental losses are present which reduce the probability of photon detection at both 780 nm and 1534 nm.
These include drift in the coupling of photons produced by the ion into the collection fiber, reductions in transmission through the second stage filtering due to drift in QFC pump frequencies ($\approx \pm 20$ MHz), and fiber butt-couplings ($\approx 50\%$ transmission) to connect each stage in a modular fashion.
Additionally, half of the ion light is of the incorrect polarization to be converted, which can be remedied by re-configuring each QFC stage into a Sagnac-type configuration \cite{ikuta2018polarization}.
With these additional losses included, we measure per-shot detection efficiencies of $1.04 \times 10^{-3}$ for 493-nm photons (APD efficiency $\approx43\%$),  $1.23 \times 10^{-4}$ for 780-nm photons over 4 hours of runtime and $6.18 \times 10^{-6}$ for 1534-nm photons over 37.5 hours of runtime.
This gives conversion efficiencies of $19.5\%$ of 493-nm photons to 780 nm and $0.66\%$ for 493-nm photons to the telecom C-band via the two concatenated conversion stages, ignoring photons lost due to incorrect polarization. 
Removal of fiber butt coupling losses would take QFC-1 close to the expected $37\%$ photon conversion efficiency.
The 1534-nm signal however, only reaches $\approx 45\%$ of its expected value given the known losses in the system. We attribute the remaining loss to factors such as drifts in pump frequency, polarization, and alignment over the experimental run time.

To demonstrate that the QFC preserves the temporal pulse shape of the photons emitted by the ion, we perform time-resolved fluorescence measurements.
We compare the temporal pulse shape of the photons at each stage of conversion in Fig.~\ref{fig:setup}(e).
Each of the photon profiles are area-normalized after background subtraction.
The overlap of the photon profiles at each color demonstrates preservation of the photon temporal shape after each stage of QFC.

We characterize the preservation of quantum statistics after the frequency conversion via a measurement of the second order intensity correlation function, $G^{(2)}(n)$ of photons emitted by the ion.
Here, $n$ represents the number of experimental cycles between photon detection events on separate detectors.
As in \cite{Siverns19convert}, we measure correlations between PMT photons and fiber-coupled photons, treating the ion itself as a beamsplitter in a Hanbury-Brown-Twiss type setup.
Three separate correlation functions are measured: $G^{(2)}_{493}(n)$, between PMT-photons and 493-nm fiber-coupled photons, $G^{(2)}_{780}(n)$, between PMT photons and 780-nm photons produced by QFC 1, and $G^{(2)}_{1534}(n)$, between PMT photons and 1534-nm photons produced by QFC 2. 
For each experiment, we time-tag and record (Picoharp 300) all detector events as well as the trigger pulses tied to each experimental cycle. 
Additionally, we use both physical and in-software gating to increase the signal-to-noise ratio of the data used for all correlation calculations.

Using the collected time-tag data, we produce histograms of ion-produced photon arrival times, relative to the 650-nm excitation pulse. 
An additional 650-nm pulse is applied 380 ns after the start of the initial excitation pulse to enable measurement of the time-dependent profile of background light present during photon extraction.
With these additional pulses and Doppler cooling, the experimental repetition rate is $\approx420$ kHz.
The result of these measurements are shown in Fig.~\ref{fig:photonShapes} for both 493-nm photons detected on the PMT and for 1534-nm photons detected on the SNDPD during the $G^{(2)}_{1534}(n)$ measurement.
In the case of PMT-photons, 650-nm light scatter and detector dark counts dominate the measured background near 1600 ns.
This is not the case for the converted 1534-nm photon profile shown in black in Fig.~\ref{fig:photonShapes}. Here, the background is attributable to detector dark counts and to Raman noise produced by the pump laser in QFC 2.
Due to the low signal-to-noise ratio of the 1534-nm photon signal, we also display a background-subtracted 1534-nm photon profile in red in Fig.~\ref{fig:photonShapes}.

The vertical blue and red dashed lines in Fig.~\ref{fig:photonShapes} indicate the software gate windows used to produce the the $G^{(2)}_{1534}(n)$ measurement, with similar gates being used for the 493-nm and 780-nm measurements. 
We calculate correlations between photons occurring in each signal window, and use the noise windows as a running measure of background in each channel.
Gate widths of 60 ns are used for the 493-nm photons detected on both the PMT and APD, as well as for the 780-nm photon data. 
To increase the signal-to-noise ratio of the 1534-nm photon signal, we use a narrower 36 ns gate.
To compensate for drifts in photon arrival time throughout the experiments (due to drifts in AOM turn on time), we reference the gate positions on all channels to the position of the PMT photon arrival time, with this position being updated every hour.

\begin{figure}[t]
\centering\includegraphics[width=\columnwidth]{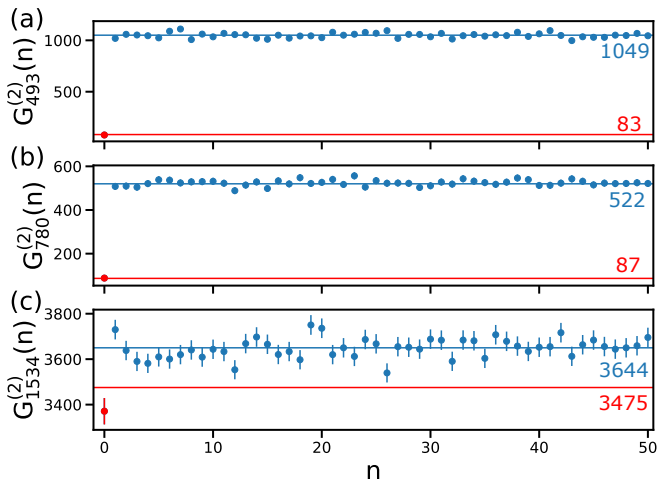}
\caption{Correlations between 493-nm PMT photons and: (a) 493-nm photons detected on the APD, (b) 780-nm photons detected on the APD, (c) 1534-nm photons detected on the SNSPD. 
Lines in each plot represent the expected number of coincidences for $n=0$ (red) and $n\neq0$ (blue), calculated using Eqns \ref{ExpectedDip} and \ref{ExpectedPeak}. 
The colored text in each plot gives the values for these lines to the nearest whole number. Error bars are statistical. Run-times: (a) 1 hr, (b) 4 hr, (c) 37.5 hr.}\label{fig:g2s}
\end{figure}

Figure \ref{fig:g2s} shows the results of all correlation measurements.
The displayed correlations are calculated as the mean between $G^{(2)}(-n)$ and $G^{(2)}(n)$ for all $n\neq 0$, using the fact that $G^{(2)}(-n) = G^{(2)}(n)$ to reduce statistical fluctuations \cite{Craddock19}. 

Theory lines in each plot of Fig. \ref{fig:g2s} represent the total number of expected correlations in the case of no quantum statistics (blue), as well as the expected noise correlation level (red), for a single photon source with the measured background noise. 
This noise correlation level includes correlations between the noise in each channel, as well as correlations between signal in one channel with noise in the other.
With $R$ photon production attempts over the course of each experiment, we calculate the background noise as

\begin{equation}
    \label{ExpectedDip}
    G^{(2)}_{theory}(0) = (C_1^S C_2^N + C_1^N C_2^S - C_1^N C_2^N)/R,
\end{equation}
\noindent where $C_x^S$ and $C_x^N$ represent the total number of counts measured in the signal and noise windows, respectively, for the $x$th photon channel.
In all cases, the dominant source of noise correlations are between PMT signal events and noise events on the APD (493-nm or 780-nm photons) or SNSPD channels (1534-nm photons).
In a similar manner, we calculate

\begin{equation}
    \label{ExpectedPeak}
    G^{(2)}_{theory}(n\neq0) = C_1^S C_2^S/R.
\end{equation}

As seen in Figure \ref{fig:g2s}, we find good agreement between the measured and expected values for both $G^{(2)}_{493}(n)$ and $G^{(2)}_{780}(n)$. 
In addition, due to the relatively high single-photon signal-to-noise ratio in these experiments (15.7 and 5.6 for the 493-nm and 780-nm photon channels respectively), we observe a large separation between correlations measured with $n=0$ and correlations measured with $n\neq0$ at these colors.
With these comparisons, we establish the quantum statistics of the 493-nm photons emitted by the ion, and show that such statistics are preserved after conversion of these photons to 780 nm.

Measuring correlations between the 493-nm PMT photons and frequency converted 1534-nm photons, we a observe a large reduction in contrast between $G^{(2)}_{1534}(0)$ and $G^{(2)}_{1534}(n\neq0)$, due to the low signal-to-noise ratio of 0.04 for the 1534-nm photon channel as measured with the SNSPD.
After the experimental runtime of $\approx37.5$ hours, we observe a 4.8 $\sigma$ separation between $G^{2}_{1534}(0)$ and  $<G^{2}_{1534}(n\neq0)>$, with $<G^{2}_{1534}(n\neq0)>$, in good agreement with the value predicted using Eqn \ref{ExpectedPeak}.
We additionally find agreement between the measured value of $3371 \pm 58$ coincidences for $G^{2}_{1534}(0)$ and our predicted value of 3475 coincidences, with under $2 \sigma$ separation between these values.
This suggests that the measured value of $G^{2}_{1534}(0)$ is wholly attributable to noise correlations as calculated by Eqn \ref{ExpectedDip}.

Practical use of this two-stage conversion scheme will require improvements to the signal-to-noise ratio.
This can be achieved by either an increase in the single photon signal or by altering the QFC pump wavelengths to reduce the amount of noise. 
The 493-nm photon probability can be increased using state-of-the-art photon collection \cite{araneda2020panopticon,shu2011efficient} or cavity-ion coupling techniques\cite{Krutyanskiy2019,SchuppIonCavity}, which have shown per-shot photon probabilities of up to 46\% into a singe-mode fiber\cite{SchuppIonCavity}, as compared to the $<1\%$ achieved in this experiment.
Additionally, QFC 1 can be improved through the use of anti-reflection coatings on the PPLN and coupling optics, which we estimate to provide a factor of 1.5 improvement over the current setup.
Finally, installing dedicated fiber connections between ion collection and the two QFC stages can remove fiber butt couplings that currently halve photon detection probability.
These improvements could increase the SNR of the 1534-nm photon channel to $>10$ without any reduction in noise.

The noise can be reduced through use of a pump laser at different wavelengths, optimized to minimize anti-Stokes Raman scattering noise  \cite{pelc2011long} at the target signal, while still converting to the telecom S- or C-band. 
For instance, a pump at $\approx 1640$ nm should reduce the noise by a factor of $\approx 10$, and convert the photon to $\approx 1515$ nm (S-band), with no changes to QFC 1 \cite{pelc2011long}.
To greatly reduce noise, the second stage conversion could instead target the telecom O-band, requiring a pump laser at $\approx$1930 nm. 
This should reduce the anti-Stokes Raman scattering noise by well over a factor of 1000.
This could serve as a good compromise between fiber transmission distance and noise reduction, while preserving the ability to integrate the system into a hybrid networking architecture. 


In conclusion, we have shown the conversion of 493-nm single photons produced by a trapped Ba$^+$ ion to the telecom C-band.
This two-stage conversion provides a pathway for C-band photons directly entangled with trapped ion ground-state qubits and also allows for potential hybrid interactions with neutral Rb-based quantum platforms.
This work is an important step in connecting trapped ion quantum systems into existing telecommunications infrastructure for long-range trapped ion based quantum networks.

\section*{Funding and Acknowledgments}
E.W. would like to acknowledge support from the National Science Foundation (grant numbers EFMA1741651, OIA2040695), and the Air Force Office of Scientific Research (grant numbers FA95501610421, FA95501810161).
Q.Q. would like to acknowledge support from the Army Research Laboratory. 

%







\bibliography{Bibliography}
\end{document}